\documentclass[a4paper]{article}

\usepackage{bm}
\usepackage{tikz,float}
\usepackage{amsmath}
\usepackage{amsfonts}

\usepackage{xcolor}
\newif\ifmathmode
\newcommand{\SA}[2]{\relax\ifmmode\mathmodetrue\else\mathmodefalse\fi\ifmathmode{\textcolor{gray}{#1}}{\textcolor{red}{#2}}\else{\textcolor{gray}{#1}}{\textcolor{orange}{---#2---}}\fi}
\newcommand{\AC}[2]{\relax\ifmmode\mathmodetrue\else\mathmodefalse\fi\ifmathmode{\textcolor{gray}{#1}}{\textcolor{purple}{#2}}\else{\textcolor{gray}{#1}}{\textcolor{blue}{---#2---}}\fi}
\title{On the viability of higher order theories}
	
\author{A. Collavini$^{1,\dagger}$ and S. Ansoldi$^{2,\ddagger}$}

\date{{\small{}$^{1}$Institute of Philosophical Studies, University of the Italian Switzerland}\\[2mm]
{\small{}$^{2}$Department of Mathematics, Computer Science, and Physics, University of Udine,\\[-1mm]
and Institute for Fundamental Physics of the Universe, Trieste, Italy}\\[5mm]}

\begin{document}

\maketitle

\begin{abstract}
	In physics, all dynamical equations that describe fundamental interactions are second order ordinary differential equations in the time derivatives. In the literature, this property is traced back to a result obtained by Ostrogradski in the mid 19th century, which is the technical basis of a \emph{no-go} theorem for higher order theories. In this work, we review the connection of symmetry properties with the order of dynamical equations, before reconsidering Ostrogradski's result. Then, we show how Ostrogradski's conclusion is reached by applying to higher order theories concepts and method that have been specifically developed for second order theories. We discuss a potential lack of consistency in this approach, to support the claim that Ostrogradski's result applies to a class of higher order theories that is nowhere representative of generic ones: we support this claim by giving an example of a higher-order Lagrangian that is asymptotically stable, but that would be unstable under Ostrogradski's criterion. We also conclude that, when considering higher order theories as fundamental, we may need to reconsider and extend the conceptual framework on which our standard treatment of second order theories is based.\\[3mm]
    \noindent{\footnotesize{}\textbf{subject}: \emph{foundations of physics}, \emph{theoretical physics}}\\
    \noindent{\footnotesize{}\textbf{keywords}: \emph{higher order theories}, \emph{Ostrogradski's instability}, \emph{symmetry and physical concepts}}\\[3mm]
    \noindent{\footnotesize{}\emph{Submitted to Philosophical Transactions of the Royal Society A}}\\
    \hrule
    \smallskip
    \noindent{\footnotesize{}%
    ${}^{\dagger}$\texttt{aaron.collavini}~\emph{at}~\texttt{hotmail.com}%
    $\quad$---$\quad$%
    ${}^{\ddagger}$\texttt{ansoldi}~\emph{at}~\texttt{fulbrightmail.org}}
\end{abstract}

\section{Introduction}
	
In physics, dynamical equations are differential equations that allow for the determination of the future evolution of a system (e.g., a system of fields), given some initial conditions/initial data and appropriate boundary conditions.
In general, there is a soft tendency to differentiate between equations describing fundamental interactions, i.e., the evolution of what we consider to be the most elementary fields, and more phenomenological inspired equations, which are often expressed in terms of other fields, derived by (or composites of) the fundamental ones.

While with the advancement of our understanding the concept of fundamental fields evolves (e.g., by re-expressing dynamical variables in terms of \emph{new, more fundamental, ones}), there is a property of the associate dynamical equations that has remained unchanged since the conception of the first modern `fundamental' dynamical equation, i.e., Newton's second law.

Indeed, Newton's second law formulates the dynamics of massive bodies in an absolute spatial and temporal background~\cite{newton1999principia} in terms of a system of \emph{second order} ordinary differential equations (ODEs). Since we have started to use Newton's second law, the presence of up to (and, including) second order time derivatives of the dynamical variables is a property that all fundamental dynamical equations of physics satisfy.

Of course, in the last couple of centuries the ontological framework on which the physical theories are based has changed a lot: not only matter is (or, seems to be~\cite{sep-quantum-field-theory}) described through fields, but the concepts of space and time themselves have been demoted from an absolute structure into a somewhat unified relativistic interpretation, which in general relativity is associated to the metric field (as written by Rovelli~\cite{Rovelli2007-ROVQG}, this new framework can be intuitively understood as ``No more fields on space-time: just fields on fields''). Then, it is even more surprising that the mathematical structure of the dynamical equations in terms of derivatives with respect to the evolution parameter has survived such deep conceptual changes. For instance, and perhaps somewhat na\"{\i}vely, this has not been the case in the passage from pre-Newtonian to Newtonian physics, as we will discuss later on. It is, therefore, natural to consider if, and why, this apparently restrictive property (the fact that equations are second order) is so deeply rooted in the structure of the fundamental dynamical equations.

In the current understanding, what we just called `restrictive' is actually often considered a necessity, apparently well explained by what is known as Ostrogradski's theorem~\cite{Woodard_2007}. Indeed, starting from some results by Ostrogradski~\cite{Ostrogradsky:1850fid}, later developed by Pais and Uhlenbeck~\cite{PaisUhlen}, a \emph{no-go} theorem has been formulated, which, roughly speaking, states that every Lagrangian theory with dynamical equations of order higher than second in the time derivatives is unstable. In this work, we aim to discuss the impact of Ostrogradski's theorem on the development of physical theories. In particular, we claim that the framework in which Ostrogradski's theorem can be applied is quite restrictive in the context of higher order theories. We support this claim by a critical review of the development of standard (i.e., second order) Lagrangian/Hamiltonian theories, with an emphasis on the relation between the degree of symmetry and the order of the fundamental equations (historically, both have increased since Aristotle). In the process, we will also emphasize some notable theories that do, actually, contain higher order derivatives. While these theories have not yet ascended to the role of fundamental theories for the description of observable physical systems, meaning that working within the framework of second order equations is not restrictive at the phenomenological level, they suggest that higher order theories might provide us effective tools to address open technical and conceptual problems.

This paper is structured as follows: in section~\ref{sec2} we develop a general argument about the role of dynamical symmetries and the definition of physical concepts. In particular, we discuss how enlarging the symmetry group results in definitions that can be related to terms with higher order derivatives of the dynamical variables. We conclude this analysis with Newton's second law, and comment on the related Lagrangian/Hamiltonian formalism, by emphasizing the tight connection with the order of the equations. As anticipated, some of the existing literature justifies how the order of the equations seems to be constrained by Ostrogradski's theorem, which we review and analyze in section~\ref{sec3}. Following the general argument given in section~\ref{sec3}, we expect that increasing the degree of symmetry of Newton's theory will result in the increase of the order of the equations. Section~\ref{sec4} shows that, indeed, a careful reconsideration of test particle dynamics in curved spacetime can be understood in terms of higher order equations when re-interpreted in a Newtonian-like way. Immediately after, we discuss how it is not challenging to find higher order equations that cannot be obtained as Euler-Lagrange equations of higher order Lagrangians: we suggest in this way that the class of theories to which the no-go theorem can be applied is not representative of generic, higher order theories. We trace back this result to the fact that Ostrogradski's theorem is built on a framework that seems tailored to the Newtonian, second order case, suggesting that its applicability to higher order theories cannot be supported at the fundamental level, and that, in any case, it lacks generality. This section ends with a technical example, in which we exhibit a higher order Lagrangian theory with fourth-order dynamical equations that, mathematically, is asymptotically stable, but that should be considered unstable according to Ostrogradski's theorem. By this example we aim to suggest the importance of developing a new technical and conceptual framework for the study of higher order theories. Finally, section~\ref{sec5} concludes the paper, by summarizing our previous analysis, and by proposing a critical discussion of related conceptual consequences.

\section{Symmetry and the definition of physical concepts\label{sec2}}

In this section we address some preliminary considerations related to the way in which dynamical equations allow us to define physical concepts. Our goal is to show that the the order of the time derivatives that appear in the dynamical equations is related to the symmetry\footnote{{In particular, as clarified by Earman in \cite{Earman1989-EARWEA-2}, we will consider the \textit{dynamical} symmetries of the laws, that should act as a constraint over the spatio-temporal ones.}} properties that we associate to our description of the systems. We will extend this analysis in~\ref{sec4}, by considering an even larger symmetry group, the group of general covariance, and showing how in a Newtonian interpretation this results in effective higher order dynamical equations for the motion of test particles. Setting this aside for the moment, we start in the next section with a somewhat na\"{\i}ve example, that we will later extend to the Aristotelian dynamics, and, finally, to Newton's law.

\subsection{A na\"{\i}ve ideal case\label{naiv}}

Let us start with an example, which is somewhat extreme compared to our current understanding of physics, and imagine that our dynamical equations relate the presence of a force to \emph{absolute positions} in space. For simplicity, we can take the space of positions to be a vector space, say ${\mathbb{R}} ^{3}$, so that positions are nothing but vectors $\bm{r} \in {\mathbb{R}} ^{3}$. We can assume that forces are also three-dimensional vectors, $\bm{f}$. In this theory we postulate the existence of a privileged point in space, $O$, so that \emph{the} position, $\bm{r}$, is an absolute concept, i.e., it only makes sense to consider positions with respect to $O$.

In this theory the `dynamical' equations for a system of $i = 1, \dots, N$ pointlike objects could naturally be (no summation implied)
\begin{equation}
	\alpha _{i} \bm{r} _{i} = \bm{f} _{i}
	,
\label{eq:zerordequ}
\end{equation}
where $\alpha _{i}$, $i = 1, \dots, N$, are appropriate proportionality constants, possibly depending on the nature of the objects. We now read these equations as follows:
\begin{quotation}
	\emph{to know if there is, and what is, the force} $\bm{f} _{i}$ \emph{acting on object} $i$ \emph{we measure the} absolute position \emph{of the object}, $\bm{r} _{i}$, \emph{and multiply it by the independently measured quantity} $\alpha _{i}$.
\end{quotation}
From the point of view that interests us here, we just wish to emphasize that the concept of force is \emph{well defined} in this theory, in the following sense: since there is no transformation, which changes our absolute determination of position and also is a \emph{symmetry transformation} (the reference point $O$ is privileged, therefore choosing a different origin for our reference is not possible without also altering the form of the `dynamical' equations), the force that we measure in our $O$-centered reference is the same force that we would measure in any other translated system,\footnote{In this work, we will talk about reference systems associated to different observers, interpreting the transformations as \textit{passive} (see, e.g., the appendix of~\cite{sep-spacetime-holearg} or \cite{pooley2020holeargument}), i.e., as coordinate transformations. This view is compatible with a relational view about space-time~\cite{sep-spacetime-theories-classical,sep-spacetime-theories,Earman1989-EARWEA-2}, as in Rovelli's interpretation of Newtonian dynamics~\cite{Rovelli2007-ROVQG}. However, our discussion is free from the active/passive debate, and from any ontological commitment about space and time, since the transformations can also be understood in active terms for a substantival manifold (see, e.g.,~\cite{Maudlin2012-MAUPOP-2}).} for the simple reason that there are \textit{no} such systems available to us in which we can still keep using~(\ref{eq:zerordequ}). We note, of course, that in principle we can measure positions with respect to a different reference point, $\bar{O}$, if so we wish. However, in this theory, before using the `dynamical equations'~(\ref{eq:zerordequ}), we must first transform the position measurement with respect to the \emph{wrong} origin $\bar{O}$ into the absolute position measurement\footnote{If we insisted on using the position with respect to $\bar{O}$, we had to compensate this choice by adding a suitable correction to the force term on the right-hand side. In this simplified setup, this corresponds to the addition of apparent forces, when in Newtonian physics the dynamics is refereed to a non-inertial reference system.} with respect to $O$ (the use of the term \emph{wrong} is emphatic, but also appropriate in the context of this theory, because it means that positions determined with respect to $\bar{O}$ are not the \emph{absolute positions} in terms of which the dynamical equations are formulated).

Paraphrasing our example in words consistent with the main topic that we are discussing in this work, we can see that our theory has an extremely low degree of symmetry with respect to space transformations (not even a translation of the origin $O$ is allowed without changing fundamental physical properties of the system, e.g., $\bm{f} _{i}$). In this theory the physical concept of force can be defined in terms of $0$-th order equations (no time derivative appears in~(\ref{eq:zerordequ})), and we can use these equations to determine the existence of a force acting on an object. This concept of force is \emph{well defined}, because all the observers that fulfill the requirement of having $O$ as an origin for position measurements (there is only one such observer) will agree on the force determinations (i.e., the definition of force is clearly non-ambiguous).

While the theory that we just described would not probably be a very effective one, it proves in an elementary setup how the degree of { dynamical} symmetry can be related to the order of the time derivatives in the dynamical equations and, in turn, to the definition of a non-ambiguous physical concept, where non-ambiguous means that \emph{all observers connected by symmetry transformations agree on the existence of the force, and on the form of the equations in which it appears}. We now wish to extend our discussion to more realistic scenarios.

\subsection{Symmetry and Aristotelian dynamics}

In the previous section we, admittedly, chose a very na\"{\i}ve example. In this subsection, we move forward to a theory that has been considered a viable description of the physical universe for a long time. In the language of the previous section, let us first consider what happens if we relax the hypothesis of a privileged origin for the determination of positions. This fact has two important implications. First, we would enlarge the symmetry group under which to seek a consistent formulation in terms of dynamical equations, by allowing the translation group to act transitively on the vector space structure, the mathematical structure behind our previous na\"{\i}ve theory. In this respect, we would be using an affine space to formulate our generalized theory.
By `realizing' this extension, we would also immediately realize how our previously defined concept of force would actually be \emph{ill defined} in this extended theory. The reason is very simple. Let us consider an observer using $O$ as a reference point for the determination of the position $\bm{r} _{i}$, which identifies the point, say, $\tilde{O}$. This observer would measure a force $\alpha _{i} \bm{r} _{i}$ acting on the object $i$. Since the origin $O$ is not absolute anymore, another observer is naturally allowed, by the symmetry of the theory, to choose a different one while still using equations~(\ref{eq:zerordequ}). For instance, an observer could choose the point $\tilde{O}$, and measure the force acting on the same object $i$ using~(\ref{eq:zerordequ}) and the position relative to $\tilde{O}$. This relative position is, of course, $\bm{0}$, therefore the observer using $\tilde{O}$ as origin would naturally conclude that there is no force acting on the object $i$. We see that,
\begin{quotation}
	\emph{by enlarging the symmetry group, which now includes the group of translations, we cannot continue to use the same dynamical equations to obtain a proper definition of a physical concept, in our case, the concept of force}.
\end{quotation}
With hindsight, this result should not by surprising. By enlarging the symmetry group, we have realized an equivalence between different descriptions of a given system that were \emph{physically} nonequivalent in the less symmetric formulation. We should upgrade our dynamical equations accordingly. A natural way to do this is to define the concept of force in terms of differences of positions, {i.e.}, in terms of velocity. This more sophisticated approach is reflected in the Aristotelian approach to dynamics.

The Aristotelian Universe, centered at the Earth, has a finite spherical shape~\cite{1983aristotle, heavensaristo, Huggett2000-HUGSFZ-2}. Among the numerous innovative ideas belonging to the philosopher's \emph{teleological} system, it is important to mention the description of the motion of objects, which is divided between natural and unnatural/violent\footnote{The world ``physics'', in fact, refers to the ``nature'' of an entity, that in the case of a moving object defines, for Aristotle, its non-accidental internal source of motion (or rest~\cite{Maudlin2012-MAUPOP-2}).}. The natural \textit{locus}-motion can be terrestrial (sublunar) or celestial (supralunar). In the former case, the four Empedoclean elements that constitute objects (i.e., fire, air, earth, water) tend to move \emph{naturally} towards their stable place, a concentric sphere in the universe. With \emph{natural motion} we mean here either the state of rest at the associated concentric sphere, or the motion towards it. Fire and air move up, while water and earth move down. In addition to the four elements, Aristotle introduces also a fifth one called \textit{aether} (or quintessence), whose stable state, conversely, is a circular motion around the center, causing part of the motion of the moon, planets, and fixed stars. What is interesting is the relationship between symmetry and the Aristotelian concept of ``force'', where by force we generically refer to the cause of deviation from the natural stable state.

In particular, we wish to investigate whether Aristotele implicitly presupposes any symmetry in the theory. Let us assume that we light a fire at some location on the earth: according to Aristotle, the element will tend to naturally move upward. If we considered the same fire at a different point on the surface of the earth (i.e., at a point translated from the previous one\footnote{In this sense, this translation is an active transformation.}), it would be natural to say, in the Aristotelian framework, that the tendency would remain the same: the fire would move up in the same way at a different location on the earth. This suggests that Aristotle is implicitly assuming a spatial symmetry under translations along the concentric spheres, whose center is the center of the universe. Such equivalence does not allow for the possibility to consider the description of motion as explicitly dependent on some absolute determination of positions, as in the na\"{\i}ve example just considered in subsection~\ref{sec2}.\ref{naiv}. Concretely, the simplest generalization of~(\ref{eq:zerordequ}) into an Aristotelian dynamical law would require first order derivatives with respect to time of the position, in modern terms the \emph{velocity}, in order to define the cause of motion (i.e., again in modern terms, the force).

Let us see if this is actually the case by considering the terrestrial motion along a concentric (sublunar\footnote{We specify that we are considering the sublunar ``corrupted'' sphere, since the argument would not run for the aether in the supralunar realm. As we have anticipated, in fact, its motion will remain uniformly circular even without any ``force''.}) circumference, as the one along the surface of the earth. Being a refined observer, Aristotle realized that as soon as a push is applied to a body, the body keeps moving. Such a motion is considered only with respect to a medium, since motion in vacuum is impossible: for Aristotle, the force contrasts the resistance offered by the medium. For example, a cart will come to rest when a horse stops pulling it, whereas a cart pulled by more horses will move faster, due to a larger ``force''~\cite{Tosca,Stinner_1994}. These considerations allow us to interpret Aristotle's dynamics as a relation between force and velocity, which in modern terms we can write
\begin{equation}
	\beta \frac{d \bm{r}}{d t} = \beta \bm{v}  = \bm{F} ,
	\label{eq:aristo}
\end{equation}
where the parameter $\beta$ may depend on the density and resistance of the medium, the mass of the object, its contact surface, and so on. Then, the above equation implies that the material objects in the sublunar sphere, if not subject to any force, remain fixed to their natural state. In the case an object is already at the corresponding concentric sphere, it remains at rest, but we can extend the same reasoning also to the other form of natural state, which is the motion towards the concentric sphere: any deviation from that vertical (radial) motion obeys equation~(\ref{eq:aristo}) as well, where the velocity $\bm{v}$ now refers to the velocity vector describing the deviation from the vertical motion with respect to the privileged reference system centered at the earth. Thus, we can conclude that the dynamical equation~(\ref{eq:aristo}) is in perfect agreement with the Aristotelian symmetry under translations. More formally, if we consider two reference systems, $O$ and $\bar{O}$, such that, say, $\bar{O}$ is individuated by $O$ through the vector $\bm{r}_0$, then the two ``Aristotelian'' observers would measure the same force, as, clearly (cf. eq.~(\ref{eq:aristo}))
\begin{equation*}
	\frac{d \bar{\bm{r}}}{d t} = \frac{d ( \bm{r} + \bm{r} _{0} )}{d t} = \frac{d \bm{r}}{d t} .
\end{equation*}
From what interests us here, we can summarize the above discussion by realizing that enlarging the symmetry group (behind the na\"{\i}ve physical theory discussed in the previous section) with the inclusion of the group of translations, naturally results in the appearance of first order derivatives with respect to time in~(\ref{eq:aristo}). Compared to the model of the previous section, the order of time derivatives has increased by one, and thanks to the invariance of the time derivative of positions under translations, the same equation~(\ref{eq:aristo}) identifies non ambiguously the concept of force, $\bm{F}$, in the following sense:
\begin{quotation}
\emph{every observer which is obtained by translating the reference system, will always agree on the presence of the force $\bm{F}$, and on its properties}.
\end{quotation}
In the following subsection we will apply similar considerations to discuss the transition from Aristotelian dynamics to Newtonian dynamics.

\subsection{Symmetry and Newtonian dynamics}

We now proceed to a second step forward in our analysis of the relation among symmetries, order of time derivatives in the dynamical equations, and non-ambiguous definitions of physical quantities. Let us consider the case of classical mechanics, i.e., Newtonian theory. In Newtonian theory, we can consider motion with respect to reference systems that are in relative motion with constant relative velocity without changing the form of the dynamical equations. Following the logic of our previous arguments, we are enlarging again the group of symmetry transformations, i.e., of the transformations that relate different reference systems resulting in the same physical description of a given dynamical system. It is known that, at the fundamental level, Newtonian physics is affected by the lack of a consistent operational definition of an inertial system, i.e. of a system in which Newton's first law holds. Our discussion is, however, unrelated to this fundamental problem, so, without restrictions, we assume that we are given at least one reference system, $O$, that \emph{is} inertial\footnote{Following Newton's bucket argument (see, e.g., \cite{sep-newton-stm, Maudlin2012-MAUPOP-2}), such an inertial system could be the one with respect to which there is no concavity on the water surface.}. Then any other system, say $\bar{O}$, related to $O$ by a {Galileian} transformation, is also an inertial one. In this context, a theory that is Galilean invariant relates the physics as measured by observers moving with constant relative velocity in space-time. This is what has been usually called the Galilean Relativity, associated with Galileo's thought experiment of the ship (interpreted in an active version~\cite{Galileo1967}), and later by Newton, in the Corollary V of the \textit{Principia}~\cite{newton1999principia}.

Let us now call $\bm{r}$ the position of an object with respect to an inertial reference system, $O$, $\bar{\bm{r}}$ the position of the same object with respect to another inertial system, $\bar{O}$, and let $\bm{V}$ be the relative constant velocity of $\bar{O}$ with respect to $O$. Then, we have
\[
    \bm{r} = \bm{\bar{r}} + \bm{V} t , 
\]
which readily translates in the Galilean law of transformation of velocities
\[
    \bm{v} = \bm{\bar{v}} + \bm{V} .
\]
We immediately see that under this enlarged symmetry group, the Galilei group, equation~(\ref{eq:aristo}) is less than desirable as a dynamical equation. In particular, the force acting on an object explicitly depends on the choice of the inertial reference system. Not only so. Let us consider a constant force, measured in an inertial reference system $O$, acting on a given body. According to~(\ref{eq:aristo}) this force would be associated to a constant velocity $\bm{v}$, also measured in the same reference system $O$. However, let us choose a reference system, $\bar{O}$, moving exactly with uniform velocity $\bm{v}$ with respect to $O$. We can freely make this choice thanks to the symmetry given by the Galilei group. With respect to $O$, the Aristotelian dynamics would predict no force acting on the object. Again, we would be in the puzzling situation that a physical concept, the presence of a force, relies heavily on the choice among different reference systems that should be equivalent in view of the symmetry properties of our dynamical theory.

As it happened before, a natural remedy to this apparent inconsistency consists in reviewing the dynamical equations, and to define the concept of force in terms of the difference of velocities, i.e., of the acceleration. This brings naturally to Newton's second law (here we restrict our attention to the case in which the mass is constant)
\begin{equation*}
	m\frac{d\bm{v}}{dt} = m\bm{a} = \bm{F},
\end{equation*}
which is the lowest order equation that allows for a non-ambiguous definition of the concept of force in a theory which is invariant under Galilei group. Indeed, and as above,
\begin{quotation}
	\emph{by enlarging the symmetry group, which now includes the group of motions with constant velocity, we cannot continue to use the Aristotelian dynamical equations to obtain a proper definition of the physical concept of force}.
\end{quotation}
Newtonian dynamics represents an extremely successful dynamical framework, and the invariance properties under the Galilei group, which are also known as the \emph{principle of relativity}, will remain central in the physical description up to the advent of general relativity. In view of the discussion in the following sections, we are also interested in the Lagrangian/Hamiltonian formulation of classical mechanics, and in discussing how such a formulation is tied to the fundamental properties of Newton's description of dynamics. Keeping this in mind, we will review basic ideas and concepts of the Lagrangian/Newtonian formulation in the following subsection.

\subsubsection{Lagrangian/Hamiltonian formulation of Newtonian dynamics}

For a system of $N$ massive particles, Newton's second law results in a system of ordinary differential equations (no summation implied)
\begin{equation}
	m _{i} \frac{d^2\bm{r} _{i}}{dt^2} = m _{i} \bm{a} _{i} = \bm{F} _{i} ,
	\qquad
	i = 1 , \dots , N
	.
	\label{eq:sysNeweqs}
\end{equation}
To solve the problem of the motion, i.e., to find the unknown functions of time $\bm{r} _{i}$, $i = 1 , \dots , N$, given the (total) force acting on each of the massive particles, $\bm{F} _{i}$, we need to solve the above system of second order, ordinary, differential equations. In the expression for the total force acting on particle $i$, $\bm{F} _{i}$, it is convenient to conceptually distinguish different contributions. In particular, one such distinctions is the separation between \emph{active} forces and \emph{reaction} forces due to the presence of constraints. Constraints are restrictions on the possible motion of the system, and they are mathematically expressed by a set of equations involving the position, velocities, and, possibly, time,\footnote{A more general definition of constraints is possible, but it is not relevant for the present discussion.}
\[
    f _{i} \left( \bm{r} _{1} , \dots , \bm{r} _{N}, \frac{d \bm{r} _{1}}{d t}, \dots , \frac{d \bm{r} _{N}}{d t} , t \right) = 0
    \qquad
    i = 1 , \dots , c
    .
\]
If the set of constraint equations is not independent, an independent subset of them can be singled out, so we will maintain the assumption of independence in what follows. In particular, the constraints are called \emph{holonomic} if they do not explicitly involve the velocities.

One convenient way not to implicitly deal with the \emph{reaction} forces due to the presence of constraints (in particular, for the case of holonomic constraints) is to reformulate the problem in the Lagrangian formalism. The Lagrangian formalism was developed almost a century after the publication of Newton's \textit{Principia} by Joseph-Louis Lagrange~\cite{lagrange1811mecanique}, and has become the theoretical framework under which all current fundamental theories of physics are studied. The way in which the Lagrangian formalism avoids explicitly the \emph{reaction} forces, due to the presence of constraints, is through a description of the dynamics in terms of a carefully chosen set of $3N - c$ coordinates (the Lagrangian or canonical/generalized coordinates). The Lagrangian approach can be consistently, and conveniently, applied to the case of time dependent constraints following the approach first proposed by Bernoulli and later by D'Alembert~\cite{d1743traite}. We briefly summarize the essential steps in what follows (see, e.g.,~\cite{goldstein:mechanics}). 

First, let us rewrite the forces in the system~(\ref{eq:sysNeweqs}) by separating the contribution of the active forces, $\bm{F} _{i} ^{(\mathrm{a})}$, from the constraints forces, $\bm{F} _{i} ^{(\mathrm{c})}$, acting on particle $i$:
\[
\bm{F} _{i} = \bm{F} _{i} ^{(\mathrm{a})} + \bm{F} _{i} ^{(\mathrm{c})} .
\]
With the above distinction, we can rewrite Newton's equations, or, better, a natural consequence of them, in terms of the vanishing of the net virtual work
\begin{equation*}
	\sum _i ^{1,N} \left[(\bm{F}_i^{(a)}+\bm{F}_i^{(c)})- m _{i} \frac{d ^{2} \bm{r} _{i}}{d t ^{2}} \right]\cdot\delta\bm{r}_i=0 .
\end{equation*}
The net virtual work (the left-hand side in the equation above) is obtained by the dot product of Newton's equations and of the virtual displacements  of the particles, $\delta \bm{r} _{i}$, which are displacements that, at any given fixed instant of time $t$, are taken by considering all the constraints and all the external forces ``frozen''. Because of their nature, virtual displacements are always perpendicular to the constraint forces, so that the virtual work of the constraints is zero, and from the last equation we obtain \textit{D'Alembert principle}
\begin{equation}
	\sum_i ^{1,N}
	\left[
	\bm{F}_i^{(a)} - m _{i} \frac{d ^{2} \bm{r} _{i}}{d t ^{2}}
	\right]
	\cdot
	\delta \bm{r}_i = 0 ,
	\label{dalprinc}
\end{equation}
which only contains the active forces. At this point, however, the virtual displacements still depend on the constraint equations. We can eliminate this dependence by choosing an appropriate set of $n = 3 N - c$ generalized coordinates, $q_i$, such that $\bm{r}_i=\bm{r}_i(q^1, \dots , q^{n} , t)$, which allows to write the virtual work as a function of the generalized force
\[
    Q_j=\sum_i ^{1,N} \bm{F}_i\cdot \frac{\partial\bm{r}_i}{\partial q^j} .
\]
In this way, the D'Alembert principle~(\ref{dalprinc}) is equivalent to the system of $n$ ordinary, second order, differential equations
\begin{equation}
	\label{form}
	\frac{d}{dt}\bigg(\frac{\partial K}{\partial \dot q^j}\bigg)-\frac{\partial K}{\partial q^j} = Q _{j} ,
	\qquad
	j = 1 , \dots , n .
\end{equation}
where $K=1/2\sum_im\,v_i^2$, $\bm{v_i}=\dot{\bm{r}_i}$, is the kinetic energy of the system,\footnote{With standard notation, an overdot denotes a derivative with respect to time.} and needs to be considered as expressed in terms of the generalized coordinates and their time derivatives. The above equations are Lagrange's equations of the first kind. If the forces $\bm{F} _{i}$ can be expressed in terms of a potential energy function $V$ as $\bm{F}_i=-\bm\nabla_i V$, then we have
\[
    Q_j=-\frac{\partial V}{\partial q^j} ,
\]
where now $V$ is now considered a function of the $q ^{j}$. In this way~(\ref{dalprinc}) finally takes the form
\begin{equation}
	\frac{d}{dt}\bigg(\frac{\partial L}{\partial \dot q^j}\bigg)-\frac{\partial L}{\partial q^j}=0,
	\qquad
	j = 1 , \dots , n,
	\label{eq:EulLageqs}
\end{equation}
where $L=K-V$ is the Lagrangian of the system. The above equations are also known as the \textit{Euler-Lagrange equations}.

Mathematically, the Lagrangian formulation is defined on the tangent bundle of the configuration space/manifold. It is standard, in classical mechanics, to reformulate the theory in an equivalent way on the cotangent bundle of the configuration space, obtaining in this way the Hamiltonian formulation of classical mechanics. In the Hamiltonian formulation, the second order system of $n$ ordinary differential equations~(\ref{eq:EulLageqs}) is replaced by an equivalent system of $2n$ differential equations of the first order, which are written in terms of the $2n$ dynamical variables $\{ q ^{i} , p _{i} \}$, $i = 1, \dots , n$ where $p _{i}$ are the conjugate momenta to the $q ^{i}$. Half of the first order equations are nothing but the relations between generalized coordinates and conjugate momenta, and are consistent with the definition of the conjugate momenta given in the Lagrangian formulation, which is
\begin{equation}
	p _{i} = \frac{\partial L}{\partial \dot{q} ^{i}},
	\qquad
	i = 1 , \dots n .
	\label{eq:conmon}
\end{equation}
In the Hamiltonian formalism, the dynamical equations are equivalent to the Hamiltonian flow on the phase space (i.e., the contangent bundle), and are written in terms of the Hamiltonian vector field, whose local expression in the $(q ^{i} , p _{i})$ chart is
\[
    \left( - \frac{\partial H}{\partial q ^{i}} , \frac{\partial H}{\partial p _{i}} \right)
,
\]
and where $H = H (q ^{i} , p _{i})$ is the Hamiltonian, or Hamilton function. The Hamiltonian is the Legendre transform of the Lagrangian
\[
H = \sum _{i=1} ^{N} p _{i} \dot{q} ^{i} - L
,
\]
where it is understood that the $\dot{q} _{i}$ are expressed in terms of the $p _{i}$ by inverting~(\ref{eq:conmon}). For future reference, we remark that, in general the $\dot{q} ^{i}$ are non-trivial functions of the $p _{i}$, which results into a non-linear expression of the Hamiltonian as a function of the conjugate momenta\footnote{Technically, the Legendre transform can be rigorously applied only if the Lagrangian is a convex functions of the $\dot{q} ^{i}$, which results in the Hamiltonian being a convex function of the $p _{i}$~\cite{bib:ArnoldVI}.}.

We conclude this essential review by referring the reader to the following sections, where we will come back to the relationship between the Lagrangian/Hamiltonian formulation and Newton's second law (in particular, the dependence of Newton's second law from second time derivatives of the dynamical variables). What we have just anticipated here is that the origin of the Lagrangian and Hamiltonian formulations is tightly bound to Newton's dynamical equations. Therefore, \emph{it must be carefully considered if it would be appropriate to apply this very same formalism to a more general situation}, a point that we will question immediately after discussing the Ostrogradski's \emph{theorem} in the next section.

\section{A review of Ostrogradski's result}
\label{sec3}
In this section we review a result obtained by Ostrogradski in the mid 19th century, which concerns theories with Lagrangians containing derivatives of order higher than the first. As a preliminary comment, let us stress that a Lagrangian containing higher order derivatives does not necessary result in higher order equations of motion. In particular, if the higher order derivatives can be rewritten as a total derivative with respect to time of any other function of the generalized coordinates and their derivatives, the dynamical equations are not higher order. For instance, if we consider a system with $n$ degrees of freedom and Lagrangian\footnote{We define here the notation
\[
    {}^{(k)}\!q ^{i} = \frac{d ^{k} q ^{i}}{d t ^{k}} 
    ,
\]
with ${}^{(0)}\!q ^{i} \equiv q ^{i}$, ${}^{(1)}\!q ^{i} \equiv \dot{q}$, and so on.}
\[
    \bar{L} (q ^{i} , \dot{q} ^{i}, \ddot{q} ^{i} , \dots , {}^{(k)}\!{q} ^{i} , t)
=
L (q ^{i} , \dot{q} ^{i})
+
\frac{d}{d t}
g (q ^{i} , \dot{q} ^{i}, \ddot{q} ^{i} , \dots , {}^{(k-1)}\!q ^{i}) ,
\]
it is easy to see that the corresponding Euler-Lagrange equations,
\begin{equation}
	\sum^{k}_{j=0}\bigg(-\frac{d}{dt}\bigg)^j\frac{\partial \bar{L}}{\partial\,{}^{(j)}\!q ^{i}}=0
	, \quad i = 1 , \dots , n ,
\label{eq:higordeullagequ}
\end{equation}
are at most second order with respect to time derivatives, as they coincide with the Euler-Lagrange equations obtained from $L$. Such, non-essential, dependence in $\bar{L}$ from the higher order derivatives is called \emph{degenerate}, and we will exclude this possibility from our framework. For generic, non-degenerate systems, the equations are instead genuinely of higher order and, generally speaking, they contain derivatives of order $2k$ if the Lagrangian contains derivatives up to order $k$ of the dynamical variables.

As an example, let us focus on a Lagrangian that contains up to second order derivatives with respect to time, i.e. let us set $k = 2$, and only one degree of freedom $q$, i.e. $n = 1$. Then, the only Euler-Lagrange equation reads
\begin{equation}
	\frac{d^2}{dt^2}\bigg(\frac{\partial L}{\partial \ddot q}\bigg)-\frac{d}{dt}\bigg(\frac{\partial L}{\partial \dot q}\bigg)+\frac{\partial L}{\partial q}=0 .
\label{eq:secordlag}
\end{equation}
The equation is fourth order if and only if the term ${\partial L}/{\partial \ddot q}$ depends on $\ddot{q}$, which corresponds to the condition of \textit{non-degeneracy}. This condition can also be written in terms of the Hessian matrix with respect to second derivatives in time of the generalized coordinates, which in our extremely simplified case reads
\begin{equation*}
	\frac{\partial^2 L}{\partial \ddot{q}^2} \neq 0.
\end{equation*}
Such condition also implies that the equations have a well-posed initial value problem with unique solution, and can be rewritten in the following form
\begin{equation*}
	{}^{(4)}\!q=\mathcal{F}(q,\dot{q},\ddot{q},\dddot{q}) .
\end{equation*}
In general, a solution $q(t)=\mathcal{Q}(t, q_0,\dot{q}_0,\ddot{q} _{0},\dddot{q}_0)$ will depend from the four initial value data $q_0$, $\dot{q}_0$, $\ddot{q} _{0}$, $\dddot{q}_0$.

It is now a non-trivial question if, and how, it could be possible to formulate a consistent Hamiltonian theory for this very simple, non-degenerate, higher order theory. What Ostrogradski proposed is to rewrite the system in terms of a set of auxiliary variables, according to the following definitions:
\begin{equation*}
	\begin{cases}
		\displaystyle Q^1\equiv q,\\[3mm]
		\displaystyle Q^2\equiv \dot{q},\\[3mm]
		\displaystyle P_1\equiv\frac{\partial L}{\partial \dot q}-\frac{d}{dt}\bigg(\frac{\partial L}{\partial \ddot q}\bigg),\\[3mm]
		\displaystyle P_2\equiv \frac{\partial L}{\partial \ddot q}.\\[3mm]
	\end{cases}
\end{equation*}
In this approach, the variables $Q ^{1}$ and $Q ^{2}$ are a set of variables that allows the Lagrangian to be written in terms of $Q ^{1}$, $Q ^{2}$, $\dot{Q} ^{1}$ and $\dot{Q} ^{2}$, and \emph{make it look} as a standard first order Lagrangian. The quantities $P _{1}$ and $P_{2}$ are, then, the canonical momenta conjugated to $Q ^{1}$ and $Q ^{2}$, and they can be understood as such, e.g., by deriving the Euler-Lagrange equations from a \emph{variational principle}, which can also be defined by \emph{mimicking} what happens for the standard Lagrangian/Hamiltonian theory that we discussed in the previous section.

From the fourth equation above, it is clear that if the system is non degenerate, then $\ddot{q}$ is a function of the three canonical coordinates $Q^1, Q^2, P_2$ only (i.e. $\ddot{q} = \mathcal{Q} (Q^1, Q^2, P_2)$), because, upon substitution, the Lagrangian only depends on $Q ^{1}$, $Q ^{2}$ and $\ddot{q} = \dot{Q} ^{2}$. Ostrogradski proposed to obtain the Hamilton function with a standard\footnote{We do not wish to enter here into the question of the convexity of this artificially constructed first order Lagrangian. We comment, however, that convexity is a sufficient condition for the uniqueness of the Legendre transform.} Legendre transform, starting from the Lagrangian obtained after performing the \emph{formal} substitutions above. In this way one obtains
\begin{equation}
	\label{ham}
	H(Q^1, Q^2, P_1, P_2)=P_1 Q^2 + P_2 \mathcal{Q}(Q^1, Q^2, P_2)-L(Q^1,Q^2, \mathcal{Q}(Q^1, Q^2, P_2)).
\end{equation}
Already at a first glance, a potential problem with the above Hamiltonian readily appears: the Hamiltonian is linear in the momentum $P _{1}$, as, one immediately sees that $P _{1}$ appears linearly and only in the first factor of the first term. This shows that the Hamiltonian~(\ref{ham}) can attain any negative value, i.e., it is unbounded from below.

The case for a general (non-degenerate) Lagrangian $L=L(q,\dot{q}, \ddot{q}, ..., {}^{(k)}\!q)$ of order $k$ describing a single degree of freedom, again $n = 1$, is obtained similarly, and contains the very same problem. In this case the $2k$ auxiliary canonical coordinates are defined as
\begin{equation*}
	\begin{cases}
		\displaystyle Q^i\equiv {}^{(i-1)}\!q, \quad i = 1 , \dots , k\\
		\displaystyle P_i\equiv\sum_{j=1}^k \bigg(-\frac{d}{dt}\bigg)^{j-i}    \frac{\partial L}{\partial\,{}^{(j)}\!q}, \quad i = 1 , \dots , k\\
	\end{cases}
\end{equation*}
and the Ostrogradski's Hamiltonian takes the form:
\begin{equation*}
	H = P_1 Q ^2 + \dots + P _{k-1} Q^{k} + P_k f(Q^1 , \dots ,Q^k,P_k)-L.
\end{equation*}
Again, the Hamiltonian is not lower bounded, and, as it is easy to realize for this case, the only assumption made is the condition of non-degeneracy. Now $k-1$ canonical variables can be the cause of the divergence towards negative energy values.

\subsection{Physical implications of {Ostrogradski's} result}

The unboundedness of the Hamiltonian discussed above is usually referred to as \emph{Ostrogradski's instability}, and we have reviewed the simplest\footnote{The derivation that we reported above is somewhat oversimplified, and not fully rigorous at the technical level. Indeed, the auxiliary system in the generalized coordinates $Q ^{i}$, $P_{i}$, $i = 1 , \dots , k$ is constrained, as, in general, $Q ^{j+1} = \dot{Q} ^{j}$, $j = 1 , \dots , k - 2$. The rigorous way to formulate the Hamiltonian theory is to apply Dirac's approach to constrained systems \cite{Dirac:1958sq}, by making sure to consider the full, closed algebra of constraints.} technical derivation of this result. It is common to consider Ostrogradski's instability as a \emph{no-go} result for generic higher order theories. It has to be stressed, however, that every mathematical issue, \emph{per se}, does not create any problem until the physical implications are analyzed.

In fact, the general argument upon which the no-go theorem is based concerns the unboundedness of the Hamiltonian. It is understood that even a Hamiltonian unbounded from below does not have to be a too serious problem for a free, i.e., non-interacting, system~\cite{Hawking_2002}: in particular it could still be just a conserved quantity, that does not have to diverge towards negative values. However, problems may arise when the system is interacting with other dynamical degrees of freedom. In this context, we report a clarifying example~\cite{Woodard_2007}, which makes more transparent the possible physical problems arising from such an unboundedness. Woodard considers a higher-derivative Lagrangian for a ``higher-order harmonic oscillator'', which is
\begin{equation*}
	L(q,\dot{q},\ddot{q})=-\frac{gm}{2\omega^2}\ddot{q}^2+\frac{m}{2}\dot{q}^2-\frac{m\omega^2}{2}q^2,
\end{equation*}
where $m$, $\omega$, $g$ are positive constants. In short, the general solution to the higher-order Euler-Lagrange equation is
\begin{equation*}
	q(t)=A_+\cos(k_+t)+B_+\sin(k_+t)+A_-\cos(k_-t)+B_-\sin(k_-t),
\end{equation*}
where the initial value constants are associated to `positive' and `negative' energy modes. This can be seen directly from the Hamiltonian of the system, which from the Ostrogradski's procedure outlined in the first part of this section is (the second line corresponds to the Hamiltonian evaluated on the general solution above):
\begin{equation}
	\label{h}
	\begin{split}
		H&=\frac{gm}{\omega^2}\dot{q}\dddot{q}-\frac{gm}{2\omega^2}\ddot{q}^2+\frac{m}{2}\dot{q}^2+\frac{m\omega^2}{2}q^2\\
		&=\frac{m}{2}\sqrt{1-4g}\,\,k_+^2(A_+^2+B_+^2)-\frac{m}{2}\sqrt{1-4g}\,\,k_-^2(A_-^2+B_-^2).
	\end{split}
\end{equation}
This example shows that:
\begin{enumerate}
	\item the problem of reaching negative energies is related to the time-dependence of the dynamical variables; i.e., it is not sufficient to check that the Hamiltonian is bounded from below for some constant configurations;
	\item the energy of the system is conserved and does not decay with time, therefore, the temporal decay of the energy cannot be considered a solution for the unboundedness of the Hamiltonian;
	\item the substantial issue concerns the mixing of positive and negative energy modes in an interacting theory; in the case of an interacting field theory, the behavior exemplified above is such that entropy favors an instantaneous decay, hence an instability, into positive and negative energy particles~\cite{Hawking_2002}; in particular
	\begin{quotation}
		``the feature that drives the instability when continuum particles are present is the vast entropy of phase space''~\cite{Woodard_2007}.
	\end{quotation}
\end{enumerate}
These arguments support the idea that the \emph{no-go} theorem related to the instability is a substantial obstacle in the development of consistent, higher-order, interacting theories. In particular, it seems to explain
\begin{quotation}
	``why Newton was right to expect that physical laws take the form
	of second order differential equations when expressed in terms of fundamental dynamical variables. Every fundamental system we have discovered since
	Newton's day has had this form. The bizarre, dubious thing would be if Newton had blundered upon a tiny subset of possible physical laws, and all our
	probing over the course of the next three centuries had never revealed the
	vastly richer possibilities. However — deep sigh — particle theorists don’t like being told something is impossible, and a definitive no-go theorem such as
	that of Ostrogradski provokes them to tortuous flights of evasion. I ought to
	know, I get called upon to referee the resulting papers often enough! No one
	has so far found a way around Ostrogradski's theorem.''~\cite{Woodard_2007}
\end{quotation}
This conclusion, however, requires a more in depth analysis. As anticipated above, numerous attempts to evade the no-go theorem have been developed (see, for instance,~\cite{Klein_2016,Motohashi_2016}), but they all involve a delicate equilibrium of constraints that, by reducing the dimensionality of the phase space, always leads to second-order equations. In the words of Swanson:
\begin{quote}
    ``To sum up: if nature is described by an interacting Lagrangian field theory with a stable vacuum, then higher than second-order equations of motions are either impossible or very special, requiring just the right interplay between constraints to eliminate the Ostrogradski's instability without reducing the dynamics to second-order laws.''~\cite{Swanson2022}
\end{quote}
It is then interesting to consider if this has to really be the only possibility, or if there can be other more substantial way around the conclusions reported above. For instance, again according to Swanson, Woodard's argument of the physical instability still relies on two main assumptions~\cite{Swanson2022}:
\begin{enumerate}
	\item the theories are \textit{field} theories;
	\item the theories admit a Lagrangian formulation.
\end{enumerate}
Now, while these observations are of a rather general nature, we think that they point in a very reasonable direction, as they suggest to investigate to which extent standard technical paradigms should be applied to higher order theories in the same way as they are in theories up to second order. We are going to perform this analysis in the following section.

\section{Viability of higher order theories\label{sec4}}

While the arguments that support Ostrogradksi result as a \emph{no-go} theorem are quite rooted into a standard understanding of the dynamics of physical systems with equations up to second order, we wish to suggest here that, at the same time, such an understanding could be \emph{just tailored for systems up to second order}. We should then wonder if it is justified to apply these methods to higher order theories.

In section~\ref{sec2}, we discussed how a consistent and non-ambiguous definition of physical concepts, as long as it is tied to the equations of motion (we used the case of force as an example), might actually require by necessity to raise the order of the time derivatives in the dynamical equations. At the same time, in some areas of physics, higher order equations show the potential to address fundamental roadblocks in extending our understanding of fundamental interactions~\cite{WeinbergQFTI}. For instance, in the context of the quantum theory of gravity, it is known that higher order Lagrangians, that can and do result in higher order equations, have better properties in connection to UV completion, and are also suggested by the renormalization of matter fields on dynamical backgrounds~\cite{Utiyama:1962sn}.

Conceptually, extending the symmetry group of a theory, as we have done in section~\ref{sec2} with, admittedly somewhat na\"{\i}ve, but also technically trivial, examples, requires us to reconsider the definition of physical quantities: non-ambiguity requires such definitions to maintain their value under a larger class of transformations, that make otherwise independent descriptions of the system, actually equivalent. We support here this idea with a natural extension of the analysis of section~\ref{sec2}, by considering the case in which we extend the symmetry group of Newtonian physics to include general coordinate transformations.

\subsection{Force on text particles and general covariance\label{SEC4A}}

In general relativity the concept of ``force due to the gravitational field'' is associated to the concept of spacetime curvature. The physical significance of curvature (gravitational force, see~\cite{hawking1975large}) is manifest in the relative acceleration between freely falling (test) particles. The appropriate technical framework to describe such relative acceleration is the study of congruences of causal curves. We will consider timelike curves in what follows, so let us consider a normalized timelike vector field in spacetime, $\bm{V}=V^{\mu}\mathbf{e}_{(\mu)}$, tangent to a timelike congruence\footnote{We denote a reference tetrad as $\mathbf{e}_{(\mu)}$, $\mu = 0 , \dots , 3$.}. Explicitly, the components of $\bm{V}$, $V^{\mu}=V^{\mu}(t, x^1, x^2, x^3)$, depend on the parameter $t$ along the integral curves. Let us consider another family of curves $\sigma(s)$ (depending on a {spacelike} parameter $s$), which are assumed a priori not to be tangent to the integral curves of $\bm{V}$. The curves are identified by a tangent vector field $\bm{Z}=\frac{\partial}{\partial s}\rvert_{\sigma(s, t)}$, interpreted as a quantification of the displacement between neighboring integral curves of $\bm{V}$. By construction, we also have that $[\bm{V},\bm{Z}]=0$, where the square brackets are the Lie brackets (Lie derivative). 

The general argument developed in~\cite{hawking1975large} is the following: we start by taking the separation vector $_\perp\!\bm{Z}$ between the curves as the projection of $\bm{Z}$ onto the subspace orthogonal to $\bm{V}$. We then take its covariant derivative
\[
    \frac{D_\perp\!\bm{Z}}{\partial t} ,
\]
and project it again perpendicularly to the congruence
\[
    \perp\!\frac{D_\perp\!\bm{Z}}{\partial t} .
\]
Finally, we take again the projected covariant derivative of this last result. What we obtain is the only \emph{natural} quantity that can describe the change in the separation of neighboring curves in the congruence in a consistent way under general covariance and the arbitrary choice of all the parametrizations involved. In particular, this quantity, which for the moment we can qualitatively describe as the relative acceleration of nearby curves of the congruence, is directly related to the Riemann tensor, thus to the curvature. For our current analysis, it is non-restrictive to assume the $\bm{V}$ congruence to be geodesic, and in this way we obtain the so-called geodesic deviation equation
\begin{equation}
	\bigg({\perp}\!\frac{D}{dt}\bigg(\perp\!\frac{D_\perp\!\bm{Z}}{dt}\bigg)\bigg)^{\lambda}=R^{\lambda}_{\,\,\kappa\beta\gamma}V^{\kappa}V^{\beta}\,_\perp\!{}Z^{\gamma}.
\label{gde}
\end{equation}
This is the technical translation of the physical idea that curvature manifests itself as \emph{relative acceleration} between freely falling particles, a quantity that can \textit{never} be made to vanish on an open set just by making an appropriate change of coordinates. As it stands, the geodesic deviation equation seems to be second order, and in a generally covariant sense it actually is. However, our goal here is to rewrite this covariant description in a way that is suitable for an interpretation in terms of the Newtonian concept of acceleration. This requires some heuristic procedure, because, in general, non-zero curvature does not allow for the existence of a global, Newtonian spacetime structure, but it will be sufficient for our goals.

In the Newtonian framework, time is not just an arbitrary parameter in terms of which worldlines/trajectories are parametrized, it is the absolute Newtonian time. However, in our current setup, geodesics of the congruence can be parametrized arbitrarily (this is why the orthogonal projection of the separation vector is required). Therefore, we need to enforce, somewhat arbitrarily, an approximate Newtonian structure to reach our goal. We proceed in the following way. First, let us choose a given geodesic, say the one with $s = s _{0}$ as a reference geodesic, so that the parameter along this geodesic will become our absolute Newtonian time. It is understood that we need to set up some Newtonian clock synchronization between the geodesic parametrized by $s _{0}$ and the one parametrized by generic values of $s$. We describe such a clock synchronization as a relationship which determines the time parameter along any other geodesic by making a specific choice for $\sigma (s , t)$. This choice will not only specify clock synchronization according to our (admittedly arbitrary, but in this sense absolute, i.e., Newtonian) choice of the time flow along our privileged geodesic identified by $s _{0}$, but will also allow to find an expression for the parameter $s$ as a function of $t$ according to this absolute synchronization. We can then realize a Newtonian interpretation by considering $s = s(t)$. This absolute clock synchronization can be set up in such a way that the corresponding, now privileged, i.e. absolute, spatial separation (written, e.g., in terms of $s$) allows us to identify $_\perp\!\bm{Z}$ with $\bm{Z}$ and $D / d t$ with $d / dt$. Moreover, if we call $\bm{r}$ the spatial separation `vector' that relates the `spatial' position described by $s$ with the reference one $s _{0}$, we have
\begin{equation*}
	\bm{Z} \sim \frac{\partial \bm{r}}{\partial s}=\frac{\partial \bm{r}}{\partial t}\frac{\partial t}{\partial s}=\frac{1}{\dot{s}(t)}\frac{\partial \bm{r}}{\partial t}.
\end{equation*}
We remember that, because of the special setup that we are considering, in terms of the absolute Newtonian time (i.e., the privileged parameter associated to the curve of the congruence identified by $s _{}0$) we can drop the projection from the previous equations, so that
\begin{equation}
	\bigg(\perp\!\frac{D}{dt}\bigg(\perp\!\frac{D_\perp\!\textbf{Z}}{dt}\bigg)\bigg) \sim \frac{1}{\dot{s}}\frac{\partial^3\bm{r}}{\partial t^3}+f(\dot{\bm{r}}, \ddot{\bm{r}}, t).
\label{eq:higordgencov}
\end{equation}
This shows, heuristically, that in fact the left-hand side of~(\ref{gde}) contains third derivatives with respect to `Newtonian' time of the `Newtonian' position of a test particle. Because of the general properties of~(\ref{gde}) this is an effect that cannot be completely removed just by making a different choice of reference system, and in this sense it represents a well-defined, i.e., non-ambiguous definition of the effect of curvature, which is the gravitational field in general relativity. Summarizing, again
\begin{quotation}
    	\emph{by enlarging the symmetry group, which now includes all general coordinate transformations (general covariance), we have increased by one the order of the time derivatives, which appear in the dynamical equations that can provide a proper definition of the physical concept of force}.
\end{quotation}
The result of this example completes the analysis that we performed in section~\ref{sec2} with a more substantial case. All the relationships that we have considered between the degree of symmetry of a system and the order of the associated dynamical equations are summarized in the following table.
\begin{center}
	\begin{tabular}{|r|c|l|}
		\hline
 		\multicolumn{3}{|c|}{\textit{Degree of dynamical symmetries and order}}\\
        \multicolumn{3}{|c|}{\textit{of the derivatives that appear in the dynamical equations}}\\
		\hline
		{\textbf{Theory}}& \textbf{Order of Symmetry} & \textbf{Order of Law} \\
		\hline
		Na\"{i}ve theory & -  & Zeroth \\
		Aristotle & Zeroth  & First \\
		Newton & First   & Second \\
		Einstein & Second & Third \\
		??? & $N$-th & $>N$-th \\
		\hline
	\end{tabular}
\end{center}
In view of this analysis, it is compelling to consider what would have happened if, before the development of general relativity, one would have had an intuition of a dynamical equation with terms of the form~(\ref{eq:higordgencov}) by only knowing the classical Lagrangian approach and Ostrogradski's theorem. Well, first, as we will discuss in the next session, the dynamics described in this higher order formulation would not have been Lagrangian, anyway\footnote{As we discussed below, for systems with only one degree of freedom, it is impossible to obtain third order dynamical equations within the standard Lagrangian formalism. This becomes possible only in appropriate interacting theories with more degrees of freedom (see, e.g.,~\cite{Motohashi_2015}).}. However, if for the moment we forget about this, and make the assumption that such an equation could have been derived from a Lagrangian theory, this would have had necessarily to be a higher order theory, and the \emph{new} equation would have probably been dismissed in view of Ostrogradski's instability.

Summarizing, this example supports again the idea that reaching higher orders in the time derivatives could be a natural process when moving to theories with larger symmetry groups. It also shows that there are concrete, and somewhat subtle, cases, in which application of Ostrogradski's theorem would result in dismissing theories that in an extended framework are absolutely viable. Keeping this in mind, in what follows we will expand on the idea that Ostrogradski's result actually applies to a very restricted class of higher order theories, which are nowhere a faithful representation of the possibly richness of higher order dynamics. We start on this path by showing, in the next subsection, that it is not hard to find higher order equations that could never be obtained within the standard Lagrangian framework. This will allow us to question the generality of this framework for higher order theories, and its applicability in the derivation of Ostrogradski's result.

\subsection{A higher order theory that is not Lagrangian in the standard sense\label{sectord}}

First, we wish to consider theories with one dynamical variable, which have third order equations, e.g., of the form
\[
\dddot{q} = f (q , \dot{q}, \ddot{q}, t) .
\]
At a closer inspection, we see that these theories cannot be obtained from a standard Lagrangian theory. It is obvious that such an equation cannot come from a first order Lagrangian, as this choice does not allow for derivatives of order higher than second. If we move to the next simple option, a non-degenerate Lagrangian, $L (q, \dot{q}, \ddot{q}, t)$ that depends on $\ddot{q}$, we immediately see that also in this case the equation above cannot be the Euler Lagrange equation for such a Lagrangian. Indeed, the Euler-Lagrange equation for a second order Lagrangian with one degree of freedom is~(\ref{eq:secordlag}). By inspecting this equation, we realize that third order derivatives cannot appear in the last term, as the Lagrangian is up to second order, and no additional time derivative appears in the last term. Of course, they can appear in the second term, but only thanks to the contribution
\[
- \frac{\partial ^{2} L}{\partial \ddot{q} \partial \dot{q}} \dddot{q} .
\]
However, if this term is non zero, this means that
\[
\frac{\partial L}{\partial \ddot{q}}
\]
is also non zero, and that it is actually just a function of $\dot{q}$ and $q$, but not of $\ddot{q}$ (otherwise ${}^{(4)}\!q$ would necessarily also appear in the dynamical equation, contrary to our assumption that it is third order). When taking the second derivative with respect to time of this contribution, the term containing $\dddot{q}$ is exactly
\[
\frac{\partial ^{2} L}{\partial \ddot{q} \partial \dot{q}} \dddot{q} ,
\]
which cancels with the one above as it is opposite to it.

We see, then, that third order equations of motion are impossible in the standard Lagrangian approach for systems with only one degree of freedom. As we anticipated, third order equations can be realized in systems with more, interacting, degrees of freedom~\cite{Motohashi_2015}. Still, the case that we considered suggests very strongly that the standard Lagrangian formalism is not general enough to include a large class of higher order theories. However, there seems to be no clear reasons why these theories must not have some fundamental relevance in physics. To further support the idea that higher order theories should not be treated and interpreted just within the standard Lagrangian framework, in the following subsection we emphasize how this framework is intimately connected to second order dynamical equations, and suggest that it is not an overstatement to consider it tailor made for this very particular case.

\subsection{The standard Lagrangian formalism as tailor made for second order equations\label{subsex:secordLagtaimad}}

Given the hint provided by the result in the previous subsection, we argue here that the standard Lagrangian approach is tailored made for second order systems. It is, then, natural to question if and why it should be plainly applied to higher order theories. While for second order theories we know \emph{by direct experience} that the Lagrangian/Hamiltonian formalism is general enough to encompass a very large class of physical theories, or, at least, all the fundamental ones that we are currently using, we have no such evidence for higher order theories. 

We will now review the basic conceptual framework behind Lagrangian theory in an oversimplified case, which contains, however, all the relevant points that we need to discuss. Let us, then, consider a system with one degree of freedom, $q$, of constant mass $m$, subject to no constraints, under the action of a conservative force given by a potential (energy) function $U (q)$. Newton's second law is simply
\begin{equation}
	m \frac{d}{d t} \left ( \frac{d q}{d t} \right) = - \frac{d U}{d q}
	.
\label{eq:Newlawsimcas}
\end{equation}
We can naturally rewrite this equation as follows
\[
	\frac{d}{d t} \left[ \frac{d}{d \dot{q}} \left( \frac{m}{2} \dot{q} ^{2} \right) \right]
	=
	- \frac{d U}{d q}
	,
\]
or, which is the same,
\[
	\frac{d}{d t} \left( \frac{\partial L}{\partial \dot{q}} \right)
	-
	\frac{\partial L}{\partial q}
	=
	0
	,
\]
where
\[
	L = L ( q , \dot{q} ) = \frac{m}{2} \dot{q} ^{2} - U (q)
	\ .
\]
This is the essence of the more sophisticated derivation that we outlined in section~\ref{sec2}, because the simple system that we are considering is not subject to any constraints, and, in particular, not subject to time dependent constraints. In this framework, the concept of energy plays a central role, and the Hamiltonian can be directly identified with the total mechanical energy of the system:
\[
	H = p \dot{q} - L = \frac{m}{2} \dot{q} ^{2} + U (q) = E
	.
\]
The reason why the concept of energy is so crucial in classical mechanics is, again, tied directly to the second order nature of Newton's law. Indeed, in our oversimplified example, we see immediately that $\dot{q}$ is an integrating factor of the second order differential equation~(\ref{eq:Newlawsimcas}). This means that
\begin{equation}
		m \ddot{q} \dot{q} = - \frac{d U}{d q} \dot{q}
		\quad \Rightarrow \quad
		\frac{d}{d t} \left( \frac{m}{2} \dot{q} ^{2} \right)
		=
		- \frac{d U}{d t}
		\quad \Rightarrow \quad
		\frac{d E}{d t} = 0
		.
\end{equation}
For a non-conservative system, again, the concept of energy enters the scene thanks to the work-energy theorem. Indeed, substituting a non-conservative force, $F (q)$, for the conservative one in the right-hand side of~(\ref{eq:Newlawsimcas}) we have
\[
		m \ddot{q} = F
		\quad \Rightarrow \quad
		m \ddot{q} \dot{q} = F \dot{q}
		\quad \Rightarrow \quad
		\frac{d}{d t} \left( \frac{m}{2} \dot{q} ^{2} \right)
		=
		F \dot{q}
		\quad \Rightarrow \quad
		K _{2} - K _{1} = \, \, \, \, \, \int _{\!\!\!\!\!\!\!\!\gamma\,\,\,\,\,1} ^{2} F dq
		,
\]
where $K = m \dot{q} ^{2} / 2$ is the kinetic energy, and the right hand side is nothing but the work done by $F$ during the motion of the system from configuration $q _{1}$ to configuration $q _{2}$ along the trajectory $\gamma$. It could now appear quite surprising if this procedure could be extended to more general systems than the one we just discussed: however, Lagrangian and Hamiltonian theories realize exactly this extension, by considerably enlarging the class of systems that can be treated in this way. As a matter of fact, this class encompasses almost every system with second order dynamical equations one could practically think of, and also includes generalized potentials that depend explicitly on the generalized velocities. It is beyond doubt that systems up to second order can find in the Lagrangian treatment a solid and consistent framework for their formulation, including systems with infinite many degrees of freedom (as in field theories).

However, it is not clear at all why the approach above should be appropriate for systems governed by higher order dynamical equations. Indeed, we wish to stress that, while oversimplified, the procedure shown above contains all the essential technical aspects that make, say, energy an appropriate physical concept for systems defined in terms of second order equations. However, the possibility to extend the same procedure to systems of higher order equations, seems not very reasonable. While there can be a restricted class of higher order systems for which an analogous procedure can successfully be carried out, this is unlikely to work for generic higher order systems. For these systems, different conceptual frameworks should also be considered. Ostrogradski's approach, however, insists on implementing the standard Lagrangian and Hamiltonian approach to such systems. As the discussion of subsection~\ref{subsex:secordLagtaimad} above shows, this necessarily, and likely severely, restricts the class of systems to which Ostrogradski's procedure can be applied in a conceptually consistent way. It is only within this class of systems that Ostrogradski's instability necessarily appears, and it is only to this class that the related \emph{no-go} theorem should be applied. For more general higher order theories, the question of whether they can be viable or not should rest on a more fundamental analysis, which is more than a routine application of techniques that have been successfully tested and validated, but only in a less general framework.

To better exemplify the above reasoning, we will now consider, explicitly, a higher order system for which a stability criterion can be given mathematically, but that would be considered unstable according to Ostrogradski's approach. We will start this analysis by first considering some general results about the stability of higher order systems of ordinary differential equations.

\subsection{Mathematical and physical stability of an higher order system}

In order to discuss the applicability of Ostrogradski's results to \emph{sufficiently general} higher order theories, an example like the one that we have discussed in subsection~\ref{SEC4A} of this section is, of course, insufficient. In this subsection, we will provide examples of higher order ordinary differential equations that satisfy a mathematical criterion of stability. In one example we will discuss equations of a form that is considered mathematically stable, but that cannot, in general, be obtained from a Lagrangian formulation. In a second example, we will exhibit an equation that is known to be mathematically stable, that can be obtained from a Lagrangian formulation, but that would be unstable in view of Ostrogradski's theorem. This second results supports the idea that Ostrogradski's theorem may be a restrictive one as far as higher order equations are concerned.

\subsubsection{Mathematical stability\label{stabode}}

Before discussing in detail the examples, we report, without derivations, some general results about the stability of higher order ordinary differential equations (ODEs). Let us start by considering linear differential equations with constant coefficients: 
\begin{equation*}
\label{ODEs}
    \bigg(a_o \frac{d^n}{dt^n}+a_1\frac{d^{n-1}}{dt^{n-1}}+...+a_{n-1}\frac{d}{dt}+a_n\bigg)y(t)=f(t).
\end{equation*}
The solutions to this equations are known if one can find the roots of the associated characteristic polynomial
\begin{equation}
\label{ahes}
    a_0x^n+a_1 x^{n-1}+...+a_{n-1}x+a_n=0.
\end{equation}
For instance, a  stability criterion can be given, that requires only the negativity of the real part of all the roots. However, Ruffini and Abel's classic result states that it is impossible to find a solution in radicals to any polynomial equation of degree higher than four~\cite{ruffini1813riflessioni, Abel_2012}. This means that, it is impossible, in general, to find the roots of the characteristic polynomial. This is why it is useful to have criteria expressed in terms of the coefficients of the ODE. One such criteria, known as the set of \emph{Routh-Hurwitz conditions}, gives the following two conditions as necessary and sufficient\footnote{See also~\cite{coppel1965stability,Bacciotti2019}.}~\cite{routh1877treatise,Hurwitz1895UeberDB}:
\begin{enumerate}
    \item $a _{0} > 0$;
    \item given the matrix
        \[
            \begin{bmatrix}
                a_1 & a_0 & 0 & 0 & 0 & 0 & \vdots & 0\\
                a_3 & a_2 & a_1 & 0 & 0 & 0 & \vdots & 0\\
                a_5 & a_4 & a_3 & a_2 & a_1& a_0 & \vdots & 0\\
                \cdots & \cdots & \cdots & \cdots & \cdots& \cdots& \ddots &\cdots\\
                0 & 0 & 0 & 0 & 0 & 0 & \vdots & a _{n}
        \end{bmatrix}
        ,
        \]
        all the $n$ principal determinants,
        \[
            \Delta_1=a_1, \quad
            \Delta_2=
            \begin{vmatrix}
                a_1 & a_0\\
                a_3 & a_2
            \end{vmatrix}, \quad \dots , \quad
            \Delta_n=
            \begin{vmatrix}
                a_1 & a_0 & 0 & \vdots & 0\\
                a_3 & a_2 & a_1 & \vdots & 0\\
                a_5 & a_4 & a_3 & \vdots & 0\\
                \cdots & \cdots & \cdots & \ddots &\cdots\\
                0 & 0 & 0 & \cdots & a _{n}
            \end{vmatrix}
        \]
        have to be positive.
\end{enumerate}
The Routh-Hurwitz criterion is a powerful tool, but it can only be applied to a restricted class of equations. It turns out that the conditions for the stability of a more general classes of higher-order equations have already been developed in the literature, after the initial work of Lyapunov~\cite{Lyapu}. Below we list some non-linear (fourth-order) prototype equations~\cite{Chin1989, Cemil1049, Wu1998, TuncC}:
\begin{align}
    &{}^{(4)}\!x\displaystyle+a_1{}^{(3)}\!x+a_2\ddot{x}+a_3\dot{x}+f(x)=0,\nonumber\\
    &{}^{(4)}\!x\displaystyle+a_1{}^{(3)}\!x+\psi(\dot{x})\ddot{x}+a_3\dot{x}+a_4 x=0,\nonumber\\
    &{}^{(4)}\!x\displaystyle+a_1{}^{(3)}\!x+f(x, \dot{x})\ddot{x}+a_3\dot{x}+f(x)=0,\nonumber\\
    &{}^{(4)}\!x\displaystyle+\psi(\ddot{x}){}^{(3)}\!x+f(x, \dot{x})\ddot{x}+g(\dot{x})+h(x)=p(t, x,\dot{x}, \ddot{x}, {}^{(3)}\!x) \label{fode} ,
\end{align}
where the $a$'s are constant coefficients. For differential equations of these forms, precise criteria for asymptotic stability can be given for some precise conditions on the coefficients~\cite{TuncC}. We will use all these results in some explicit examples that can be found in the following subsection.

\subsubsection{Two case studies}

\paragraph{Linear fourth order equations.} As we have anticipated in the previous section, the mathematical stability of the solutions of linear, higher-order equations, can be studied with the Routh-Hurwitz conditions. Let us restrict our attention to the lowest higher-order equations that can be obtained from a standard Lagrangian approach for a system with one degree of freedom. These equations are given by~(\ref{eq:higordeullagequ}), with $k = 4$. Explicitly, we can write
\begin{equation}
\begin{split}
    &\bigg(\frac{\partial^2 L}{\partial\ddot{q}^2}\bigg)q^{(4)}+\bigg(2\frac{\partial^3 L}{\partial q\partial\ddot{q}^2}\dot{q}+2\frac{\partial^3 L}{\partial\dot{q}\partial\ddot{q}^2}\ddot{q}+\frac{\partial^3L}{\partial\ddot{q}^3}\dddot{q}\bigg)\dddot{q}+\\
    &\qquad+\bigg(2\frac{\partial^3 L}{\partial q\partial\dot{q}\partial\ddot{q}}\dot{q}+\frac{\partial^3 L}{\partial \dot{q}^2\partial\ddot{q}}\ddot{q}+\frac{\partial^2L}{\partial q\partial\ddot{q}}-\frac{\partial^2L}{\partial\dot{q}^2}\bigg)\ddot{q}+\bigg(\frac{\partial^3 L}{\partial {q}^2\partial\ddot{q}}\dot{q}-\frac{\partial^2 L}{\partial {q}\partial\dot{q}}\bigg)\dot{q}+\frac{\partial L}{\partial q}=0.
\label{eq:4theullagexp}
\end{split}
\end{equation}
For a fourth order system, the Routh-Hurwiz condition can be rewritten as
\[
    a _{0}, a _{1} , a _{2} , a _{3} , a _{4} > 0
    \quad \mathrm{and} \quad
    a _{1} a _{2} - a _{0} a _{3} > 0
    \quad \mathrm{and} \quad
    a _{1} a _{2} a _{3} - a _{1} ^{2} a _{4} - a _{0} a _{3} ^{2} > 0
    .
\]
By direct comparison with~(\ref{eq:4theullagexp}) we see that it is not possible to obtain an equation with all non-zero constant coefficient, and we conclude that fourth-order equations that are stable according to the Routh-Hurwitz criterion cannot be obtained from a Lagrangian description. This supports the idea that there are technical and conceptual features in the Lagrangian approach that are tailored to second order systems, and it may not be appropriate to follow Ostrogradski's approach, in applying this second order framework to the reformulation of a higher order Lagrangian in terms of auxiliary variables.

\paragraph{Nonlinear fourth order Euler-Lagrange equations.} We conclude this section with an additional example, in which we check if it is possible to obtain some of the non-linear equations discussed above from a Lagrangian theory. In particular, let us consider the Lagrangian
\begin{equation}
\label{ostrocontro?}
    L(q,\dot{q},\ddot{q})= A\ddot{q}^2+B\dot{q}\ddot{q}^2+C\dot{q}^2+Dq^2\ddot{q}+E q\dot{q}+Fq^2+Gq .
\end{equation}
The corresponding equations of motion are
\begin{equation}
    A{}^{(4)}\!q+(2B\ddot{q})\dddot{q}+(2Dq-C)\ddot{q}+(D\dot{q}^2+B\dot{q})+(2Fq+G)=0
\end{equation}
and we see that this form falls within the general form~(\ref{fode}), as long as we make the following identifications (we also set $A = 1$ for simplicity)
\[
    \psi(\ddot{q})=2B\ddot{q} , \quad
    f(q,\dot{q})=2Dq-C , \quad
    g(\dot{q})=D\dot{q}^2+B\dot{q} , \quad
    h(q)=2Fq+G , \quad
    p(t, q,\dot{q}, \ddot{q}, \dddot{q})=0.
\]
According to Ostrogradski's theorem, such a system would be unstable for all values of the constants $B$, $C$, $D$, $F$, $G$, but, as discussed in the previous section, it is possible to find values of these constants for which the solutions are asymptotically stable~\cite{TuncC}. This raises the question if the definition of stability implied by Ostrogradski's approach is appropriate for higher-order systems, for which, as we discussed, the concept of energy defined through the Hamiltonian reformulation in terms of Ostrogradski's auxiliary variables might not be an appropriate one.

In any case, these examples support the conclusion that, rather than excluding from consideration higher order theories in view of the \emph{no-go} theorem extrapolated from Ostrogradski's result, it might be appropriate, instead, to reconsider the formulation and physical interpretation of theories with higher order dynamical equations.

\section{Conclusion and Discussion\label{sec5}}

The development of physical theories is never as linear as it may seem. The birth of a new framework is a challenging process, which consists in several attempts, reformulations and a lot of critical thinking. A well known example is the reconciliation between the constancy of the speed of light and the Galilean relativity, that ended in a deep, technical and conceptual, rethinking about space and time. Very often, when scientists face new phenomena that do not fit existing frameworks, the only available tools that they can use are those that are already known. This is probably an unavoidable step in the development of new concepts and ideas, but at some point some discontinuity is needed. In this process, no-go theorems often represent a very difficult hurdle to overcome.

In this work, we have considered higher order theories, i.e., theories that contain derivatives with respect to the evolution parameter (time) of order higher than second. As of today, observations/experiments do not seem to require such a generalized framework, with some notable exceptions. For instance, in quantum gravity renormalizability properties of higher order theories are more desirable than those of canonical approaches applied to general relativity (for an early example, see~\cite{PhysRevD.16.953}). Also at the classical level, modified gravity theories with higher order equations, like $f(R)$ gravity~\cite{RevModPhys.82.451,DeFelice:2010aj}, have desirable properties, which make them viable (although not exclusive) candidates to tackle specific problems, for instance in cosmology. In other situations, like Horndeski's theory~\cite{Horndeski:1974wa}, the opposite approach is followed, i.e., theories have been developed that are the most general containing up to second order equations. In this context, Ostrogradski's theorem is the result that, more than anything else, seems to constrain higher order theories, to the point that it has raised to the status of a no-go theorem.

In this work we have critically reviewed not so much the Ostrogradski's result, which is clear both at the technical and conceptual level, but the no-go theorem status that has acquired over time. There are technical reasons behind this analysis, e.g., the fact that Ostrogradski's approach applies methods that have been developed for second order theories to higher order ones. The Lagrangian/Hamiltonian approach falls among these methods. Additionally, Ostrogradski's approach applies it by rewriting higher order theories in terms of auxiliary variables, so that formally they look like a second order theory, which, of course, they are not. While Ostrogradski's approach is the most well known one, it is noteworthy to realize that other approaches have also been proposed. A notable example is the work by Masterov~\cite{Masterov_2016}, in turn based on the work of Bolonek and Kosinski~\cite{bolonek2005hamiltonianstructurespaisuhlenbeckoscillator}, who derives a $\mathcal{N}=2$ supersymmetric Pais-Uhlenbeck oscillator that has \emph{both} a stable ground state and bounded energy spectrum.

Here, we intend to challenge the impact that the no-go theorem based on Ostrogradski's result might have at the conceptual level, and we discussed several points that should be kept in mind in this respect.

First, the increase in the order of time derivatives in the dynamical equations can be justified in terms of symmetry arguments. By considering dynamical symmetries, we argued that when enlarging the symmetry group we may actually have to increase the order of fundamental equations, on which the definition of fundamental physical concepts can be based. This can be intuitively justified, because enlarging the symmetry group renders equivalent those descriptions of the physical system that were distinguishable under a lower degree of symmetry. This is a strong motivation \emph{not to ban} higher order theories in principle.

Second, after reviewing the basic ideas and definitions behind the Lagrangian and Hamiltonian formulations of classical mechanics, we emphasized how they are tightly bound, technically, to the fact that Newton's second law is second order in the time derivatives. We intuitively supported this claim, by emphasizing that there are wide classes of higher order differential equations that cannot be obtained from an extension of the standard Lagrangian/Hamiltonian framework to higher orders. There is, in principle, no reason why these differential equations should not be viable models of physical systems.

Third, supported by these considerations, we reviewed the main ideas behind Ostrogradski's instability, and emphasized how they are strongly tied to an approach that, somewhat arbitrarily, makes use of a first order effective description, to which the standard Lagrangian/Hamiltonian treatment is then applied. In the more general context of higher order equations, it is not clear at all why the stability of a higher order system should be judged based on this effective Hamiltonian formulation.

Fourth, we showed that it is possible to find mathematically stable higher order differential equations that, (i) do not fit in the standard Lagrangian/Hamiltonian formulation, and (ii) would be unstable under Ostrogradski's analysis: there is, however, no reason why we should assume that such equations cannot be made sense of in a larger framework that is more suited to discuss higher order equations (for instance, jet spaces could be a reasonable mathematical framework to include higher order theories~\cite{Crampin_Sarlet_Cantrijn_1986}).

With our, mostly conceptual, analysis, we wish to suggest how it might be appropriate to investigate the possibility to develop new technical and conceptual frameworks for the physical application and interpretation of higher order theory, and how the implicit acceptance of a no-go theorem could be detrimental to the development of our physical understanding. Indeed, while each of the elements that concur to such a no-go theorem are, individually, important and fundamental ingredients in almost every physical theory that is currently accepted, to us their interplay in the formulation no-go theorem associated to Ostrogradski could create an implicit barrier to the development of our physical understanding of the universe. Therefore, we would like to conclude this paper with a very simple example that conceptually supports this point of view.

Let us consider a second order equation, and assume that it describes well our experiments/observations of physical systems. For simplicity, let this equation be of the standard form
\begin{equation}
    \frac{d ^{2} q}{d t ^{2}} = - \frac{d V}{d q}
    ,
\label{eq:sim__q}
\end{equation}
where $q = q (t)$ is the only degree of freedom of the system, and $V = V (q)$. This theory, as we know, has a consistent Lagrangian/Hamiltonian formulation, e.g., one possible Lagrangian is
\[
    L (q , \dot{q} ) = \frac{\dot{q} ^{2}}{2} - V (q)
    .
\]
Let us now write our theory in terms of an auxiliary variable $\phi (t)$, such that $q (t) = \dot{\phi} (t)$. Then the dynamical equation would be replaced by the higher order equation
\begin{equation}
    \frac{d ^{3} \phi}{d t ^{3}} = - \frac{d \bar{V}}{d \dot{\phi}}
    ,
\label{eq:simphi}
\end{equation}
where $\bar{V} = \bar{V} (\dot{\phi}) = V (q(\dot{\phi}))$. Correspondingly, we could write a function
\[
    \bar{L} ( \ddot{\phi} , \dot{\phi} ) = \frac{\ddot{\phi} ^{2}}{2} - \bar{V} (\dot{\phi})
    =
    L \left( \frac{d}{dt} q(\dot{\phi}) , q (\dot{\phi}) \right)
    .
\]
It is natural at this point to comment that there is certainly a way to make the theory described by~(\ref{eq:sim__q}) equivalent to the theory described by~(\ref{eq:simphi}). As a first comment, note, however, that the equation of motion coming from $\bar{L}$ differs from~(\ref{eq:simphi}), although it can surely be made equivalent to it under the same conditions under which (\ref{eq:sim__q}) is equivalent to~(\ref{eq:simphi}). However, our point is somewhat more elaborated.

Let us now imagine that our equation is just our best approximation to a deeper fundamental theory governed by the following equation:
\begin{equation}
    \frac{d ^{3} \phi}{d t ^{3}} = - \frac{d \bar{V}}{d \dot{\phi}} + \Psi (\phi, \dot{\phi}, \ddot{\phi}, t)
    .
\label{eq:simphifun}
\end{equation}
For instance, our technology could be inadequate to observe the effects due to the presence of the function $\Psi$ in the dynamical equation~(\ref{eq:simphifun}), which could be heavily suppressed compared to the effects caused by $\bar{V}$. Having no evidence of the physics associated to $\Psi$, very likely our best theory would be the one given by~(\ref{eq:sim__q}). Even if, accidentally (and very unlikely), we would actually `discover' equation~(\ref{eq:simphi}) (in all respects, a very lucky coincidence), we could immediately realize that, theoretically (and maybe phenomenologically, under appropriate circumstances) it would be natural to actually work with the simpler, second order, version~(\ref{eq:sim__q}). 

Given these premises, one should seriously wonder what would happen to our future understanding of physics, if we did not have just our ``fundamental'', or maybe so believed, equation~(\ref{eq:sim__q}), but if we had it paired with the no-go theorem inspired by Ostrogradski's instability.

\medskip

\section*{\small{}Acknowledgments}
\vskip -2 mm
{\small{}One of us, SA, would like to thank the Theoretical Astrophysics Group of Kyoto University for hospitality during the conceptualization of the ideas presented in this work. SA also warmly acknowledges fruitful discussion and brainstorming with Takahiro Tanaka and Hideki Ishihara.}


\end{document}